\documentstyle[aps,pre,epsfig]{revtex}
\begin{document}
\draft
\title{\bf Parameter estimation in spatially extended systems: \\ 
The Karhunen-L\'{o}eve and Galerkin multiple shooting approach\footnote{This preprint published in Phys. Rev E \bf{64} 056222 (2001)}}
\author{Anandamohan Ghosh, V. Ravi Kumar\footnote{e-mail for correspondence: ravi@che.ncl.res.in}, B. D. Kulkarni}
\address{Chemical Engineering Division, National Chemical
Laboratory, Pune 411 008, India}
\date\today
\maketitle
\begin{abstract}
Parameter estimation for spatiotemporal dynamics for
coupled map lattices and continuous time domain systems
is shown using a combination of multiple shooting,
Karhunen-Lo\'{e}ve decomposition and Galerkin's projection methodologies.
The resulting advantages in estimating parameters have been studied and
discussed for chaotic and turbulent dynamics using small amounts of data
from subsystems, availability of only scalar and noisy time
series data, effects of space-time parameter variations,
and in the presence of multiple time-scales.
\end{abstract}

\section{Introduction}
In general, extensive systems exhibiting complex spatiotemporal dynamics
including chaos may be studied as processes involving reaction-diffusion 
and convective mechanisms \cite{Cr93}.
Analysis of the dynamics of these systems is not an easy task because 
of the large attractor dimensions involved.
It would be desirable to develop ways of studying spatiotemporal systems 
using reduced model descriptions in conjunction with subsystem dynamics
especially when it is known that extensive scaling relationships in 
dynamics  exist as a function of subsystem size \cite{Gr89,Ba91,Pa98,St99a}.
Methods developed for low-dimensional systems may then become applicable 
with the concomitant advantage of simplifying the data  
requirements for studying the spatiotemporal dynamics.
In this paper, we show a profitable use of this approach for parameter 
estimation with reduced model descriptions of spatiotemporal systems 
and which uses subsystem data for the characterization.
The low-dimensional models can be obtained by projecting the  
governing equations onto relevant modes obtained by Karhunen-Lo\'{e}ve (KL) 
decomposition \cite{Si87} along with Galerkin projection \cite{Ho96}.  
Efficient ways such as multiple shooting boundary value algorithms 
\cite{Ba83,Ti98,Ti00} that are known to curtail error propagation
for low-dimensional chaotic and noisy data for continuous systems \cite{Ba92},
may be then reformulated, as shown here, to estimating true parameter 
values using reduced models for the spatiotemporal dynamics. 

There has been a great deal of interest in forecasting spatiotemporal 
time series and model identification using polynomial and mixed 
functions \cite{Ba99,Pa00,Vo98}.
Short-term prediction of spatiotemporal dynamics by reconstruction of 
local states \cite{Pa00}, and KL decomposition using empirical basis
functions by training amplitude coefficients using genetic 
algorithms for optimization \cite{Lo00} have been studied and assessed.
Rather than phase space reconstruction models, identification using some 
knowledge of the system structure along with nonlinear parametric regression 
\cite{Vo98} has also been used for modeling spatially extended systems. 
In this context, the attempt here is to demonstrate the potential of
a different approach that uses a multiple shooting algorithm
for parameter and state variable estimations in analyzing 
spatiotemporal behavior.

In Sec.~\ref{II}, the algorithm for
spatiotemporal model parameter 
estimation by a combination of Karhunen-Lo\'{e}ve decomposition,  
Galerkin's projection and Multiple Shooting (KLGMS) methodology is 
presented for spatiotemporal systems described by  coupled
map lattices \cite{Ka93,Wi95}
and partial differential equations.
Illustrative examples using the KLGMS approach are presented in Sec.~\ref{III} 
for a single 
variable coupled map lattice (CML) that possesses
the basic reaction-diffusion and convection 
mechanisms that give rise to complex patterns including spatiotemporal 
chaos and convective turbulence.
The use of the methodology in characterizing the model from subsystem 
data using limited number of snapshots is shown 
and the adaptability of the method in inhomogeneous model \cite{Me00} 
identification  using perturbation strategies is also discussed.
The formalism is then applied in Sec.~\ref{IV} to an 
autocatalytic reaction-diffusion system
described by multivariable partial differential equations (PDE)
and exhibiting spatiotemporal chaos \cite{Ma96}. 
Here we develop the method further for use in stringent situations 
when only scalar and noisy data in a single variable from subsystems 
is available for parameter estimation.

\section{KLGMS approach}
\label {II}
Given snapshots of spatiotemporal data, $u^{(i)}(n,j)$, for variables
$i=1,2,\cdots$ at discrete times $(n=1,2,\cdots,M)$ and 
spatial nodes $(j=1,2,\cdots,L)$,
we may obtain the fluctuating components $v^{(i)}(n,j)$ as
\begin{eqnarray}
v^{(i)}(n,j) = u^{(i)}(n,j) - \bar{u}^{(i)}(j) \label{fluc}
\end{eqnarray}
where 
\begin{eqnarray}
\bar{u}^{(i)}(j) = \frac{1}{M} \sum_{n=1}^{M} u^{(i)}(n,j) \label{avg} 
\end{eqnarray}
represent temporal averages for the $M$ number of snapshots considered.
The KL decomposition assumes $v^{(i)}(n,j)$ may be expanded in a 
separable form as
\begin{eqnarray}
v^{(i)}(n,j) = \sum_{k=1}^{M} a^{(i)}_k(n) \phi^{(i)}_k(j), \label{kle}
\end{eqnarray}
where by truncating the index $k$ to an optimum value, say $N$, it becomes 
possible to reconstruct the $v^{(i)}(n,j)$ to a required accuracy.
In Eq.~(\ref{kle}),  the $a^{(i)}_k(n)$ are time-dependent coefficients
while the $\phi^{(i)}_k(j)$ are spatial basis functions, satisfying 
the orthonormality condition
\begin{equation}
(\phi^{(i)}_l, \phi^{(i)}_m) = \delta_{lm} \label{orth}
\end {equation}
with the inner product defined as
\begin{eqnarray}
(\phi^{(i)}_l, \phi^{(i)}_m) = \sum_{j=1}^L \phi^{(i)}_l(j)
\phi^{(i)}_m(j).  \label{inp}
\end{eqnarray}
The spatial basis functions can be empirical basis functions
obtained from the spatial correlation matrix \cite{Si87},
Fourier modes, Chebyshev or other sets of orthogonal
polynomials, wavelets, etc. \cite{Ho96,Da92,Wi99}.
In particular, to obtain empirical basis functions
the KL decomposition of the data $v^{(i)}(n,j)$ [Eq.~\ref{kle}]
is carried out in an optimal fashion~\cite{Ho96} such 
that the obtained $\phi_k$ are eigenfunctions that 
maximize the associated eigenvalues 
$\lambda_k = \langle (\phi_k,v)^2\rangle$.  
Here $(\cdot,\cdot)$ denote the inner product as defined in
Eq.~(\ref{inp}) while $\langle \cdot \rangle$ represents an
averaging procedure that commutes with the inner product \cite{Ho96}. 
A factor $\eta_N=\sum_{l=1}^{N} \lambda_l / \sum_{k=1}^{M} \lambda_k$
is a measure of the energy content for increasing mode index $k$
and decides an index $N < M$ for which the series in Eq.~(\ref{kle}) may be
truncated. 
More often the spatial domain is much too large and
estimation of eigenfunctions $\phi^{(i)}_k(j)$ from a spatial
correlation matrix can become quite involved. 
The method of snapshots~\cite{Si87} helps to get over this practical
difficulty by
using instead the temporal correlation matrix $C_{lm} = (1/M)
\sum_{j=1}^{L} v(l,j)v(m,j)$ as the kernel and reduces the KL decomposition 
to solving a standard eigenvalue problem of the  form 
$C_{lm}\Phi_m = \lambda \Phi_l$ with $\Phi$ being the required 
set of eigenfunctions $\{ \phi^{(i)}_k(j) \}$. 
From the $\phi^{(i)}_k(j)$ the respective time-dependent 
coefficients $a^{(i)}_k(n)$ may be obtained from
\begin{equation}
b^{(i)}_k(n) = a^{(i)}_k(n) = (v^{(i)}(n,j),\phi^{(i)}_k(j)) \label{aemp}
\end{equation}
and a full description of the dynamics
as a series expansion [Eq.~(\ref{kle})] becomes available.
Note the introduction of notation $b^{(i)}_k(n)$
in Eq.~(\ref{aemp}) [for $a^{(i)}_k(n)$] is to convey
that the time-dependent coefficients have
been obtained solely from transformed snapshot data $v^{(i)}(n,j)$
[Eq.~(\ref{fluc})]. The known values of
$b^{(i)}_k(n)$ may, therefore, be used as time-series for model
characterization and parameter estimation studies.

Our next step is to obtain a simpler model of the
spatiotemporal system that can be used for
the purposes of parameter estimation. Such a reduced description may be
derived by  Galerkin's projection of the model in conjunction with
the KL expansion [Eq.~(\ref{kle})] and we shall discuss the
steps involved in this procedure separately for discrete CML type
systems and for continuous systems modeled by PDEs.  

A CML model with discrete space index $j$ and time $n$ may be written
in a general form as
\begin{eqnarray}
u^{(i)}(n+1,j)
= g(u^{(1)}(n,j+j^{\prime}),u^{(2)}(n,j+j^{\prime}),\cdots,\mu), \label{gcml}
\end{eqnarray}
where, $j^{\prime} = \cdots,-2,-1,0,1,2,\cdots$ and
$g$ any function with $\mu = \{\mu_1,\mu_2,\cdots,\mu_p\}$
the model parameters.
The boundary conditions for Eq.~(\ref{gcml}) are generally dependent
on the example
being studied and in Sec.~\ref{III} we show cases involving periodic
and open flow (as in convection) boundary conditions.
In the KL-Galerkin's method we minimize the residual
by forcing the projection of the model on the subspace of truncated basis
functions $\phi^{(i)}_k(j)$ to be zero at all time $n$ and obtain
$N < M$ KL-Galerkin equations for the CML model described by Eq.~(\ref{gcml})
as 
\begin{eqnarray}
a^{(i)}_k(n+1) = \sum_{j=1}^{L} \big[ g\big(
\sum_{l=1}^{N} a^{(1)}_l(n) \phi^{(1)}_l(j) + \bar{u}^{(1)}(j),
\cdots,\mu \big) -
\bar{u}^{(i)}(j)\big]\phi^{(i)}_k(j) \label{gal}
\end{eqnarray}
where the use of the orthonormal property
of the basis functions [Eq.~(\ref{orth})] allows considerable
simplification. The initial conditions for solving the map [Eq.~(\ref{gal})]
can be obtained by forcing the initial residual to have zero 
projection on the space of basis functions, {\it i.e.},
$a^{(i)}_k(0) = \sum_{j=1}^{L} v^{(i)}(0,j) \phi^{(i)}_k(j).$
 
For continuous systems in both  space $x$ and time $t$,
the mathematical model can be written in general form as PDE's
with appropriate boundary conditions, {\it viz.},
\begin{eqnarray}
\dot{u}^{(i)}(t,x) = h(u^{(1)}(t,x),u^{(2)}(t,x),\cdots,\mu) \label{gpde}
\end{eqnarray}
and the corresponding
KL-Galerkin equations by projection obtained as
\begin{eqnarray}
\dot{a}^{(i)}_k(t) = \int h \big(
\sum_{l=1}^N a^{(1)}_l(t) \phi^{(1)}_l(x) + \bar{u}^{(1)}(x),
\cdots, \mu \big) \phi^{(i)}_k(x) dx. \label{galc}
\end{eqnarray}
Here again $N<M$ equations form a simpler and reduced
model incorporating system parameters $\mu$.
Note that the summation in Eq.~(\ref{inp}) for discrete systems becomes an 
integral for continuous ones, {\it i.e.},
$(\phi^{(i)}_l, \phi^{(i)}_m) = \int_{0}^{1} \phi^{(i)}_l(x)
\phi^{(i)}_m(x) dx$ 
with the $(\cdot,\cdot)$ now denoting the usual ${\bf L}^2([0,1])$
inner product space \cite{Ho96} defined in spatial domain $0 \leq x \leq 1$. 
The initial conditions for solving Eq.~(\ref{galc}) may again
be independently obtained by 
$a^{(i)}_l(0) = \int v^{(i)}(0,x) \phi^{(i)}_l(x) dx$. 

We next discuss the use of the simpler KL-Galerkin models  
in the time-dependent coefficients $a^{(i)}_k(n)$
[{\it i.e.}, [Eq.~(\ref{gal}) for CML or Eq.~(\ref{galc}) for a 
continuous system], along with the
known coefficient values $b^{(i)}_k(n)$ obtained from the
data by [Eq.~(\ref{aemp})] for the estimation of 
parameters $\mu$ for the respective 
spatiotemporal dynamics [{\it i.e.}, Eq.~(\ref{gcml}) or Eq.~(\ref{gpde})].
We show this is possible by using an
effective algorithm in parameter estimation for low-dimensional
nonlinear dynamical systems, {\it viz.}, the multiple shooting
algorithm formulated as a multipoint boundary value problem with
nonlinear constraints for optimization~\cite{Ba83,Ba92}. The algorithm
curtails error propagation observed in chaotic dynamics and offers
advantages in terms of number of data points required, negating
effects of noise, handling missing data situations, parallelization 
and stopping at local
minima during optimization~\cite{Ti98,Ti00}.

For the CML model [Eq.~(\ref{gcml})], the observations in 
the discrete time interval $[n_1, n_M]$ may be
chosen to form a grid for $M$ multiple shooting points at $n_1<n_2
<\cdots<n_l<\cdots<n_M$ modes forming $(M-1)$ sets of initial value
problems in the form of [Eq.~(\ref{gal})]
for each of the shooting nodes $n_l$ with $1 \leq l \leq M-1$. 
That is, on considering $k=1,2, \cdots,N$
spatial basis modes for $N<M$, we obtain for a $i^{\rm th}$ variable
$N(M-1)$ maps to be solved
\begin{eqnarray}
a^{(i)}_k(n+1) = {\cal G}(a^{(i)}_k(n),\phi^{(i)}_k(j),
\bar{u}^{(i)}(j),\mu) ~~~~~~~ 
a^{(i)}_k(n_l) = s^{(i)}_k(l),\label{ivpd}
\end{eqnarray}
for an incremental time-step $n \rightarrow n+1$. Here, 
$s^{(i)}_k(l)$ denotes the value of the $i^{\rm th}$ variable
at the $l^{\rm th}$ shooting-point 
for $k^{\rm th}$  basis mode and initial guesses for solving $N(M-1)$ maps of
Eq.~(\ref{ivpd}) are taken to be the known values of
$b^{(i)}_k(l)$ [Eq.~(\ref{aemp})].
It may be noted that
for the system governed by PDEs and data available at shooting points 
$\tau_1 < \tau_2 \cdots < \tau_l < \cdots < \tau_M$ 
and that are monitored at time 
$n \Delta t$, the corresponding set of $N(M-1)$ initial value problems
may be written as
\begin{eqnarray}
\dot{a}^{(i)}_k(t) = {\cal H}(a^{(i)}_k(t),\phi^{(i)}_k(x),
\bar{u}^{(i)}(x),\mu), ~~~~~~a^{(i)}_k(\tau_l) = s^{(i)}_k(l).\label{ivpc}
\end{eqnarray}

We can construct an augmented vector of 
initial values $s$ and parameters $\mu$ 
for either model Eq.~(\ref{gcml}) or Eq.~(\ref{gpde}), {\it viz.},
\begin{eqnarray}
{\bf z} = (s^{(1)}_1(1),s^{(1)}_1(2),\cdots,s^{(1)}_1(M),
s^{(2)}_1(1),s^{(2)}_1(2),\cdots, \mu_1,\mu_2,\cdots,\mu_p) \label{optvar}
\end{eqnarray}
and attempt to minimize a least square cost function ${\cal L}_2({\bf z})$
of the form
\begin{eqnarray}
{\cal L}_2({\bf z}) = \sum_{i=1,2,\cdots}\sum_{k=1}^N \sum_{n=1}^M
\frac{1}{\sigma^{(i)}_{kn}}
[b^{(i)}_k(n) - {\cal F}^{(i)}_k (a(n),{\bf \mu})]^2, \label{cost}
\end{eqnarray}
where, ${\cal F}^{(i)}_k$ is a function relating components of ${\cal G}$ in
Eq.~(\ref{ivpd}) or ${\cal H}$ in Eq.~(\ref{ivpc}) 
and comparing to the known $b^{(i)}_k(n)$ with $\sigma^{(i)}_{kn}$
the square of the standard deviation.
The minimization in Eq.~(\ref{cost}) is carried out subject to satisfying
\begin{eqnarray}
a^{(i)}(n_{l+1})-s^{(i)}(l+1) \rightarrow 0 \label{equality}
\end{eqnarray}
so that the trajectories in the coefficients $a^{(i)}(n_{l+1})$
become continuous.
Alternatively stated, by identifying
${\bf y}_1({\bf z}) = r(s^{(i)}_1(1), s^{(i)}_2(2),
\cdots, s^{(i)}_N(M), \mu_1,\mu_2,\cdots,\mu_p)$ 
and ${\bf y}_2({\bf z}) = a^{(i)}(n_{l+1}) - s^{(i)}(l+1)$
we obtain, a standard nonlinear minimization problem~\cite{Po78} of the type 
\begin{eqnarray}
\min_{\bf z} \{ \parallel {\bf y}_1({\bf z}) \parallel_2^2~|~
{\bf y}_2({\bf z}) \rightarrow 0  \} \label{min1}
\end{eqnarray}
where the minimization of ${\bf y}_1({\bf z})$ corresponds to 
minimizing the cost function Eq.~(\ref{cost}) while that for
${\bf y}_2({\bf z})$ implies satisfying the constraints imposed by 
Eq.~(\ref{equality}).
The  minimization of Eq.~(\ref{min1}) can be carried out by
starting with initial guess values
${\bf z}^{(0)}$ and iterating for ${\bf z}$ using
${\bf z}^{(q+1)} = {\bf z}^{(q)} + \omega^{(q)} \Delta {\bf z}^{(q)}$
where, $\omega^{(q)} \in [0,1]$ are damping factors.
In doing so corrections to the augmented vector ${\bf z}$, {\it viz.},
$\Delta {\bf z}^{(q)}$ are obtained by solving the linearized problem
\begin{eqnarray}
\min_{\bf z} \{
\parallel {\bf y}_1({\bf z}^{(q)}) + \frac{\partial {\bf y}_1({\bf z}^{(q)})}
{\partial {\bf z}} \Delta{\bf z}^{(q)}\parallel^2_2~~|~~
{\bf y}_2({\bf z}^{(q)}) 
+ \frac{\partial {\bf y}_2({\bf z}^{(q)})}{\partial {\bf z}}
\Delta{\bf z}^{(q)} \rightarrow 0 \} \label{min2}
\end{eqnarray}
The above Eq.~(\ref{min2}) may be solved by a suitable nonlinear optimization
technique in the optimization variables,
${\bf z}$
[Eq.~(\ref{optvar}] for arbitrary guess values for the parameters $\mu$ 
and initial states $s^{(i)}_k(l)=b^{(i)}_k(l)$.
In the coding of the above KLGMS approach,
we have employed the successive quadratic programming
algorithm \cite{Sc86} coupled with numerical differentiation for the
sensitivity matrices. 
For illustration, we have retained simplicity in the cost function 
[Eq.~(\ref{cost})] but more effective functionals~\cite{Mc99}
may be adopted in the optimizing for the parameters $\mu$.
It is to be noted that the methodology also allows optimizing for the 
$s^{(i)}_k(l)$ even when some values of $b^{(i)}_k(l)$
are initially not available and arising due to missing snapshot data in  
say some $i^{\rm th}$ variable. The illustrative examples presented in
Sec.~\ref{III} and Sec.~\ref{IV} bring out further the advantages
offered by KLGMS approach when snapshot data availability
is limited and possibly noise contaminated.

\section{Parameter estimation using KLGMS for CML}
\label{III}
\begin{figure}
\mbox{\epsfig{figure=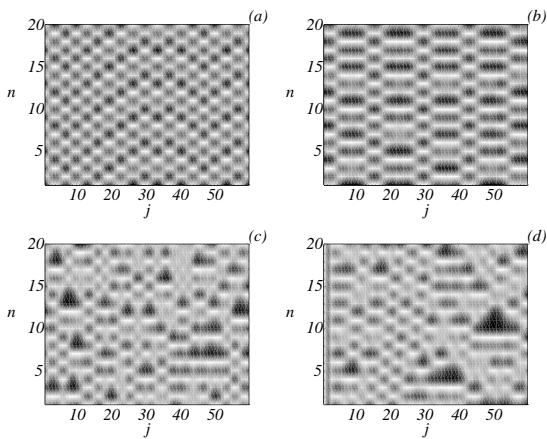,height=58mm}} \caption{Evolved
spatiotemporal data $u(n,j)$ for the CML ($j$ spatial grid with
$L=60$; $M=20$ snapshots. {\it(a)}~Weak chaos ($F=1.73$,
$D_d=0.4$, $D_c=0.0$); {\it(b)}~Traveling wave ($F=1.5$, $D_d=0.5$,
$D_c=0.0$); {\it(c)}~Fully developed chaos ($F=2.0$, $D_d=0.4$,
$D_c=0.0$); {\it(d)}~Convective turbulence ($F=2.0$, $D_d=0.4$,
$D_c=0.3$). } \label{chobi1}
\end{figure}

Because of their computational simplicity, CMLs are a popular and
convenient paradigm for studying fully developed turbulence
\cite{Ka93,Wi95,Hi99}, chaos \cite{Ka89},  and pattern formation \cite{Cr93}
in systems. A CML model is a discrete space-time system with
continuous state space and studies the effects of local nonlinear
reaction dynamics, the coupling arising from diffusion due to
state space gradients as well as convective effects by asymmetric
coupling\cite{Wi95}. Here, we consider a CML involving a single
spatial dimension and incorporating these mechanisms as
\begin{eqnarray}
u(n+1,j) = (1 -D_d - D_c)f(u(n,j))  + D_c f(u(n,j-1))\nonumber\\
+ \frac{D_d}{2} \big[ f(u(n,j+1)) + f(u(n,j-1)) \big] , \label{cml}
\end{eqnarray}
where, $u(n,j),~j=1,2,\cdots,L$ is the state of the variable
located at site $j$ at time $n$ for  a lattice of size $L$, $D_d$
the nearest neighbor diffusive coupling strength, and $D_c$
denoting the asymmetric coupling constant.  
This being a single variable system we suppress the index $(i)$ in this Section.
For $D_c=0$ the system represents a reaction-diffusion system while 
for $D_c \neq 0$ mimics one with convective effects included. 
We assume the reaction dynamics on the lattice sites is governed by the
nonlinear logistic function $f(u) = 1 - F u^2, \label{map}$ where,
$F$ is the nonlinearity parameter. Thus, depending on the
parameter values for $F$, $D_d$ and $D_c$, a variety of dynamical patterns 
may be observed in Eq.~(\ref{cml}) and characterized as in \cite{Wi95}. 
We bring out the methodology for estimating parameters for selected
dynamics covering a broad range of complexity, {\it viz.},
{\it(a)}~weak chaos;  {\it(b)}~traveling wave; {\it(c)}~fully developed chaos;
and {\it(d)}~convective turbulence. Spatiotemporal data for the
different cases are obtained by evolving Eq.~(\ref{cml}). 
All the sites are given random initial conditions at
$n = 0$ and snapshots are stored after eliminating
initial transients. Cases~{\it (a,b,c)} are evolved with periodic
boundary conditions, {\it i.e.,} $u(n,1) = u(n,L)$ while for the
convective case {\it(d)} the left boundary is assumed fixed, {\it
i.e.,} $u(n,1)=1$, with the right boundary open.
The gray-scale images of the spatiotemporal data with the parameter
values yielding the data for a lattice size of $L = 60$
and for $M = 20$ snapshots, is shown
in Fig.~\ref{chobi1}. In studies involving subsystems, only the
data corresponding to the evolution of the chosen subsystems are stored.

We obtain a KL decomposition for the spatiotemporal data
$v(n,j)$ and Table~\ref{TableI} shows the corresponding eigenvalues
$\lambda_k$, and the energy content $\eta_k$, for the data shown
in Fig.~\ref{chobi1} {\it (a-d)}.  The results show that for the CML
exhibiting weak chaos and travelling wave, a smaller number of
basis modes $N=3$ and $N=5$, respectively, are required to capture
and reconstruct $99 \%$ of the data. For the more complex
patterns, {\it viz.}, fully developed chaos and convective
turbulence the number of basis modes significantly rise to $15$
for $\approx 99 \%$ and $19$ for $\approx 100 \%$ accuracy.

Studies with KL-Galerkin Eq.~(\ref{gal}) for the CML
Eq.~(\ref{cml}) using the KLGMS
approach did accurately and simultaneously estimate the unknown
parameters ($F, D_d, D_c$) from a few snapshots of the data.
The results of convergence for arbitrary and different initial guesses
for the parameters shown in Table~\ref{TableII}  
for the {\it(a)}~weakly chaotic,
{\it(b)}~travelling wave and {\it(c)}~fully developed chaos cases. The
robustness is seen when parameters were successfully estimated
even for noisy spatiotemporal data sets [Table~\ref{TableII}]
obtained by additive noise $\hat{u}(n,j) = u(n,j) + \eta$ with
Gaussian distribution noise $\eta \in N[0,\varepsilon^2]$. The
strength of the noise level used was determined by
$\sigma_{noise}/\sigma_{data}$ and chosen to be 0.01.
It may be observed that noise in the data enters through
the ``coefficient trajectories" $b_k(l)$ that are obtained by the
convolutions of the fluctuating data $v(n,j) = u(n,j) -{\bar u}(j)$
with the basis functions $\phi^{(i)}_k(j)$ {\it via,} Eq.~(\ref{aemp}).
Note that although the functional form of the CML in
the form of Eq.~(\ref{cml}) is single dimensional and single
variable the procedure may be extended to situations involving
multivariable mappings Eq.~(\ref{gcml}) and higher spatial dimensions.
The effects of considering higher spatial dimensions do not change
the methodology because the KL expansion yields two or
three-dimensional spatial basis functions $\phi(i,j,k)$ but the
Galerkin equation still retains the mapping form of Eq.~(\ref{gal})
in the time-dependent coefficients $a_k(n)$.
Applications of the method to systems with multivariable coupling and
scalar data are shown in the Sec.~\ref{IV} studying KLGMS
for continuous time systems. For brevity the
results obtained with CMLs on these aspects are not presented.

\begin{figure}
\mbox{\epsfig{figure=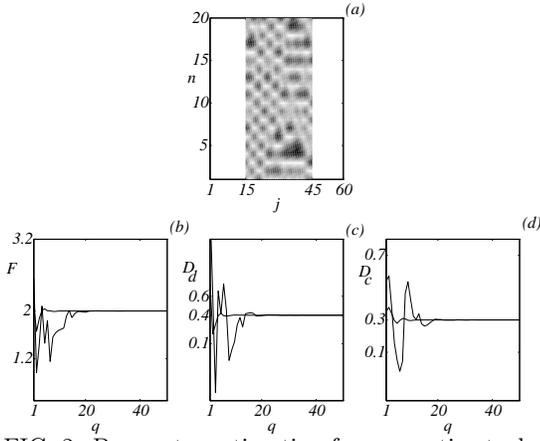,height=58mm}} \caption{Parameter
estimation for convective turbulence.
{\it(a)} Subsystem data for the central 31 lattice sites;
{\it (b,c,d)} Simultaneous convergence to parameter estimates for $F$,
$D_d$ and $D_c$ for arbitrary initial guesses (shown as {\it y}-axis labels)
as iterations $q$ proceed for minimizing
the least square functional.} \label{chobi2}
\end{figure}

The presence of scaling relationships in Lyapunov exponents as a function
of subsystem size have been studied \cite{Ba91,Pa98,St99a}.
For KL decomposition modes, using the spatial correlation matrix, a linear relationship
in KL dimension \cite{Zo97} has also been seen.
Our studies for subsystem scaling with the temporal
correlation matrix, $C_{lm}$, showed some interesting features.
We observe that ${\cal D}_{T} = {\rm max} \{N: \eta_N \leq f\}$ required to
capture a fraction $f$ of the total variance showed scaling behavior
after an optimum subsystem size before saturation.
The saturation occurs either due to the dynamics being not complicated
enough to warrant all modes to be included as a function of subsystem size
or alternatively when the dynamics is sufficiently complex that
all KL modes (limited by the number of snapshots $M$) are required.
Therefore, depending on the complexity of the pattern
and number of snapshots, $M$, an optimum subsystem size exists
beyond that only system features can be extracted reliably.
The feasibility of estimating parameters by relaxing the need for data
from the entire spatial domain was then considered.
Thus, on computing ${\cal D}_{T}$ for the convective CML data
[Fig.~\ref{chobi1}{\it(d)}] as a function of subsystem size $j$
for $L=80$ we observed that beyond $j=30$ there
is linear scaling and this determines the optimum subsystem size.
For this subsystem size even with a lower number of modes($\approx
N=15$), parameter values could be estimated, while for larger
subsystem size all KL modes need to be considered.
Figure~\ref{chobi2}{\it(a)} shows the subsystem data for the central 31
lattice sites and used for parameter estimation purposes for
$N=15$. The accurate convergence of the estimated parameters $F$,
$D_d$ and $D_c$ with search iterations is shown in Fig.~\ref{chobi2}{\it (b-d)}
and reported as the homogeneous case~~{\it(a)} in Table~\ref{TableIII}.
These studies suggest that when reliability of data is poor from certain
regions, considerable information may be gained by using only authentic
data available from other subsystems in the spatial domain.

\begin{figure}
\mbox{\epsfig{figure=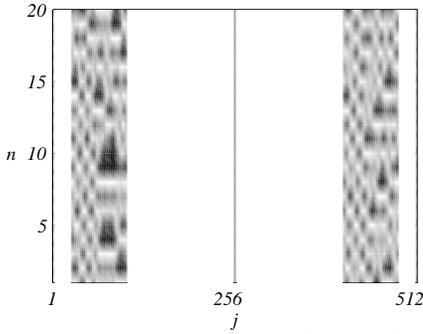,height=45mm}} \caption{Data from the
left ($F=2.0$) and the right ($F=1.9$) subsystems for the
inhomogeneous CML. The vertical line at $j=256$ marks the
boundary; Other parameter values $D_d=0.4$, $D_c=0.3$.}
\label{chobi3}
\end{figure}

A number of real situations have inhomogeneous  distribution of
parameter values in space and/or slowly varying in time domain.
Studies in this context for parameter estimation were carried out
and the analysis of a simple example is discussed here.
We evolve a CML such that the sites in the left half ({\it i.e.}, $1 \leq j
\leq 256$) have $F=2.0$ while the right half ({\it i.e.}, $257
\leq j \leq 512$) evolve data with $F=1.9$ for $L=512$.
Subsystem data from each half [Fig.~\ref{chobi3}] was used for parameter
estimation. Since the local dynamics propagate in space, the data
obtained from both subsystems had composite features
leading to inconsistent and
unreliable parameter estimates.
To overcome this difficulty we recorded data immediately after a
giving a perturbation at time $n$ ({\it i.e.}, noise of strength 0.01)
to the variable $u(n,j)$
and then carried out KLGMS parameter estimation
for each of the subsystems (left and right).
The results presented in Table~\ref{TableIII} cases {\it b,c} show that
parameter estimation is now possible.
Studies were also carried out for
situations modeling $F$ as a slowly varying parameter in time.
The need to record subsystem data at optimum time gaps was
found necessary to monitor the slow parametric changes.
In real situations, repeated parameter estimations at sufficient time
intervals can help in establishing relationships in the nature of
parametric variations and this can considerably aid system analysis.

\section{Parameter estimation using KLGMS for reaction-diffusion system}
\label{IV}
A basic problem in studying spatially extended dynamical systems
is the quantitative comparison of experimental data with models
based on partial differential equations. For example, in the study
of pattern forming systems, the theoretical models usually take the
form of reaction-diffusion equations that have been studied both
theoretically and experimentally
\cite{Cr93,Ku84,Fi85}. For our study of parameter
estimation we shall illustrate the methodology for a prototype
reaction-diffusion model where one chemical species grows
autocatalytically on another species \cite{Gr90,Ma96}. This
model is a simplification of the model of glycolysis proposed by
Selkov \cite{Se68} and it follows the reaction mechanism $U + 2V
\rightarrow 3V$; $V \rightarrow P$ with a continuous supply of the
reactant $U$ and removal of product $P$. The model has been
extensively studied from the point-of-view of pattern formation
and comparisons with features observed in experimental data have
also been attempted \cite{Le94}.

The reaction-diffusion mechanism yields a two variable PDE model
involving concentrations $u^{(1)}(t,x)$, $u^{(2)}(t,x)$ of $U$, $V$,
respectively, and for a spatially
one dimensional system, we obtain:
\begin{eqnarray}
\frac{\partial u^{(1)}(t,x)}{\partial t} &=& D_u \nabla ^2 u^{(1)}(t,x)
- u^{(1)}(t,x)(u^{(2)}(t,x))^2 + f[1 - u^{(1)}(t,x)] \nonumber\\
\frac{\partial u^{(2)}(t,x)}{\partial t} &=& D_v \nabla ^2 u^{(2)}(t,x)
+ u^{(1)}(t,x)(u^{(2)}(t,x))^2 - [f + k]u^{(2)}(t,x). \label{gs}
\end{eqnarray}
Here, $D_u$ and $D_v$ are the diffusion coefficients of species $U$ and $V$,
with parameters $f$ and $k$ related to the flow of
reactant into the system and the kinetic rate constant. The
parameters $f$,$k$ form a pair of bifurcation parameters that may be varied to
obtain a host of spatiotemporal Turing patterns for unequal diffusion
coefficients of the chemical species as seen in \cite{Ma96}.

\begin{figure}
\mbox{\epsfig{figure=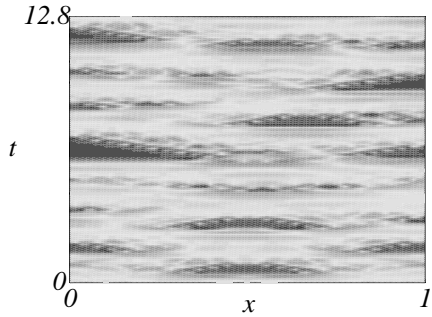,height=45mm}}
\caption{Spatiotemporal data for the variable $u^{(1)}(t,x)$ in
the autocatalytic reaction-diffusion system  with parameter values
$f=0.029$, $k=0.0535$, $D_u=0.00002$, $D_v=0.0001$ with spatial
length $L=1$ spanning $160$ spatial sites and $M=128$ snapshots
recorded at a time step  $\Delta t = 0.1$ is shown.} \label{chobi4}
\end{figure}

In our study, we consider the situation corresponding to system exhibiting
spatiotemporal chaos Fig.~\ref{chobi4}.
as studied in \cite{Ma96}.
For obtaining the spatiotemporal data $u^{(1)}(t,x), u^{(2)}(t,x)$,
Eq.~(\ref{gs}) is solved
numerically with Euler discretization in the spatial domain,
with spatial length $L=1$ spanning $160$ spatial sites and $M=40$
snapshots are stored at a time step $\Delta t=0.1$ and
with periodic boundary conditions 
$u^{(1)}(t,0)=u^{(1)}(t,L)$ and $u^{(2)}(t,0)=u^{(2)}(t,L)$ imposed.
The initial conditions correspond to the stationary solution
$u^{(1)}(0,x)=1$ and $u^{(2)}(0,x)=0$
except for a few central sites that are  given a random
perturbation to break the symmetry.

Here we will also consider situations where only scalar data 
in a single variable $u^{(1)}(t,x)$ is monitored.  
Because the data in the $u^{(2)}(t,x)$ is not
available we need to use basis functions other than empirical. 
In the present study, we choose to exemplify KLGMS using Fourier 
basis functions defined as
\begin{eqnarray}
\phi^{(i)}_k(x) = \sqrt{2} \sin(2 \pi k x) \label{fbas}
\end{eqnarray}
with temporal coefficients obtained by
\begin{eqnarray}
b^{(i)}_k(t)=a^{(i)}_k(t) = \int_0^L v^{(i)}(t,x) \phi^{(i)}_k(x) dx
\label{ab}
\end{eqnarray}
and use the $b^{(i)}_k(t)$ as observables in evaluating
the least square functional in Eq.(\ref{cost}).
For the model Eq.~(\ref{gs}) the KL Galerkin projection 
equations for the time-dependent coefficients, {\it i.e.,} Eq.~(\ref{galc}),
for modes $k=1,2,\cdots,N$
can be written as (suppressing $(x)$ and $(t)$):
\begin{eqnarray}
\dot{a}^{(1)}_i = \int_0^L [ D_u \nabla^2
(\sum_{k=1}^N a^{(1)}_k \phi^{(1)}_k + \bar{u}^{(1)})
- (\sum_{k=1}^N a^{(1)}_k \phi^{(1)}_k + \bar{u}^{(1)})
(\sum_{k=1}^N a^{(2)}_k \phi^{(2)}_k + \bar{u}^{(2)})^2 \nonumber \\
+ f (1 - \sum_{k=1}^N a^{(1)}_k \phi^{(1)}_k
- \bar{u}^{(1)}) ] \phi^{(1)}_i dx  \nonumber
\end{eqnarray}
\begin{eqnarray}
\dot{a}^{(2)}_i = \int_0^L [ D_v \nabla^2
(\sum_{k=1}^N a^{(2)}_k \phi^{(2)}_k + \bar{u}^{(2)})
+ (\sum_{k=1}^N a^{(1)}_k \phi^{(1)}_k + \bar{u}^{(1)})
(\sum_{k=1}^N a^{(2)}_k \phi^{(2)}_k + \bar{u}^{(2)})^2  \nonumber\\
- (f+k) (\sum_{k=1}^N a^{(2)}_k \phi^{(2)}_k + \bar{u}^{(2)}) ]
\phi^{(2)}_i dx   \label{galgs}
\end{eqnarray}
and  a reduced  $N<M$ set of ODEs solved by integrating using the initial
conditions discussed for Eq.~(\ref{galc}).

Our studies with the set of Galerkin equations Eqs.~(\ref{galgs}) with
KLGMS for estimating system parameters using the spatiotemporally
chaotic data (Fig.~\ref{chobi4}) showed two interesting features 
described below.
First, accurate parameter estimation of the diffusion coefficients
($D_u$, $D_v$) did not particularly depend on the choice of ($k, f$)
when initial transient data were chosen as snapshots with
the diffusion mechanism playing a significant role.
It was also observed that similar results in $(k,f)$
were obtained using
snapshots after giving a perturbation to the system state $u^{(1)}(t,x)$ 
at any time $t$.
The second feature was that having evaluated ($D_u$, $D_v$)
in the above fashion the other two
parameters ($f,k$) could  be successfully estimated
using post-transient data.
These observations suggest that diffusion rates and reaction rates occur at
differing time-scales and  clearly point to the need for suitable data
sampling strategies.
It may be noted that the values of diffusion coefficients employed here
lie in typical ranges.
The multiple time-scale features discussed above may therefore be expected
to be frequently present in the dynamics of spatiotemporal systems.
Any methodology seeking model identification would
need to consider this relevant aspect for parameter estimation.

Without any ambiguity, we discuss other features of the KLGMS with
reference to evaluating $f$ and $k$ from monitored post-transient
data. The KL decomposition of the data set using Fourier modes for
$40$ snapshots showed that a single Fourier basis mode could
reconstruct the data snapshots accurately ($>99.8$\%). The results of
parameter estimation with this single mode considered showed that
accurate convergence was consistently possible even when the data
was corrupted with noise of the order of $5$\% and are summarized
in Table~\ref{TableIV} case~{\it a}. For the present reaction-diffusion
system we have observed that the use of basis functions with
the known Fourier form allows tolerance for higher noise levels when
compared to empirical basis functions (using correlation
matrices). A more practical problem arises in multivariable
systems when only one dynamical variable is monitored. We assume
that $u^{(2)}(t,x)$ is not monitored and assign initial guesses
for the temporal coefficients $a^{(2)}(t)=0.2$ and
$\bar{u}^{(2)}=0$ for the multiple shooting algorithm. The least
square functional Eq.~(\ref{cost}) and equality constraints are
suitably modified so as to take into account only terms in
variables  $u^{(1)}(t,x)$. Results of the study presented in
Table~\ref{TableIV} case~{\it b} showing accurate parameter estimation is
again possible for both $f,k$ although with a small decrease in
noise tolerance. It may be seen that the parameter estimation of
$k$ present only in the $u^{(2)}$ equation of the PDE model
Eq.~(\ref{gs}) is also possible. Importantly, we have recovered
the unmonitored variable $u^{(2)}(t,x)$ using Eq.~(\ref{kle}) and
estimated the values of $a^{(2)}(t)$ by multiple shooting.

\begin{figure}
\mbox{\epsfig{figure=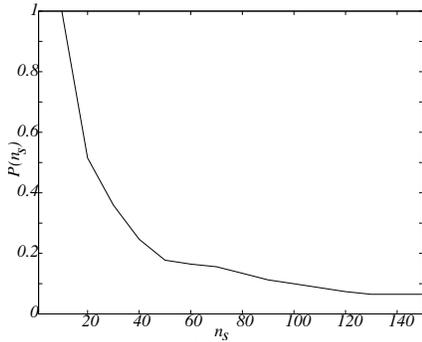,height=45mm}}
\caption{Power $P(n_s)$ in the first mode of the temporal coefficients,
normalized to the maximum, is plotted as a function of
subsystem size $n_s$.}
\label{chobi5}
\end{figure}

Similar to the studies using CML we attempted to evaluate parameters using
subsystem data with only scalar variable data in $u^{(2)}(t,x)$ available.
An indication of the optimum subsystem size in this
study using Fourier basis functions was suggested on evaluating
the normalized power $P(n_s) = \int [a^{(1)}_1(t)]^2_{n_s} dt$
as a function of the subsystem size $n_s$ and is shown in Fig.~\ref{chobi5}.
The results indicate a near saturation beyond $n_s=80$.
The results of KLGMS carried out with subsystem scalar data available
only in $u^{(1)}$ and with noise (Table~\ref{TableIV} case~{\it c})  shows that
parameter estimation within reasonable error bounds is still possible.

\section{Conclusion}
\label{V}
The results obtained using the KLGMS show that this basic framework has the
necessary robustness for parameter estimation for spatiotemporal dynamics.
We exemplify the methodology by simultaneously estimating all
parameters of a CML and a reaction-diffusion system.
Importantly, for complex dynamics and noise in the data we show that  accurate
parameter estimates are possible even from small data samples obtained
from subsystems of optimal size.
We show ways of adapting the methodology for inhomogeneous situations
when parameters vary in space and time and by using transient data soon
after perturbing the system dynamics.
The usefulness of this strategy especially when multiple time-scales
are present in the system dynamics has been discussed.
The algorithm can be extended to situations when only scalar data is available
and has the capability to recover the dynamics of the unmonitored variable.
The study presented here should also help in the analysis and design of
experiments for spatiotemporal systems that are often costly and
difficult to perform.

\acknowledgments
Authors wish to thank Unilever Research, Port Sunlight, UK, for financial
assistance in carrying out the work.

\begin{table}
\caption{Significance of KL modes in CML.} \label{TableI}
\begin{tabular}{llccc}
Case& & Mode no. $k$ & $\lambda_k$ & $\eta_k$ \\
\cline{1-5}
{\it(a)}& Weakly  & 1 & 10.5467 & 0.9408 \\
&chaotic & 2 & 0.3639  & 0.9733 \\
&  & 3 & 0.2862  & 0.9988 \\
\cline{1-5}
{\it(b)}&Traveling   & 1 & 13.2207 & 0.9337 \\
&wave & 2 & 0.3712  & 0.9600 \\
&   & 5 & 0.1098  & 0.9957 \\
\cline{1-5}
{\it(c)}&Fully    & 1  & 3.8130 & 0.2690 \\
&chaotic & 15 & 0.0661 & 0.9912 \\
&    & 19 & 0.0217 & 0.9999 \\
\cline{1-5}
{\it(d)}&Convective   & 1  & 4.0071 & 0.2950 \\
&turbulence & 15 & 0.0458 & 0.9929 \\
&    & 19 & 0.0106 & 0.9999
\end{tabular}
\end{table}

\begin{table}
\caption{Parameter estimation for the CML with varying dynamics.
Error bounds for arbitrary initial guesses are shown.} \label{TableII}
\begin{tabular}{llcc}
Case & & $F$ & $D_d$\\
\cline{1-4}
{\it(a)}& Weakly chaotic         & $1.73 \pm 0.01$ & $0.40 \pm 0.02$ \\
&with noise   & $1.74 \pm 0.03$ & $0.39 \pm 0.02$ \\
{\it(b)}& Traveling wave        & $1.50 \pm 0.01$ & $0.50 \pm 0.01$ \\
&with noise  & $1.54 \pm 0.03$ & $0.53 \pm 0.04$ \\
{\it(c)}& Fully chaotic       & $1.99 \pm 0.01$ & $0.40 \pm 0.01$ \\
&with noise & $2.03 \pm 0.04$ & $0.38 \pm 0.03$
\end{tabular}
\end{table}

\begin{table}
\caption{Parameter estimation from subsystem CML data for convective
turbulence. Error bounds for arbitrary initial guesses are shown.}
\label{TableIII}
\begin{tabular}{lcccc}
Case&& $F$ & $D_d$ & $D_c$\\
\cline{1-5}
{\it(a)}&Homogeneous  & $1.99 \pm 0.02$ & $0.41 \pm 0.03$ & $0.32 \pm 0.03$ \\
{\it(b)}&Inhomogeneous left & $1.99 \pm 0.03$ & $0.39 \pm 0.02$ & $0.29 \pm 0.04$ \\
{\it(c)}&Inhomogeneous right & $1.88 \pm 0.03$ & $0.42 \pm 0.04$ & $0.28
\pm 0.04$\
\end{tabular}
\end{table}

\begin{table}
\caption{Parameter estimation for the autocatalytic reaction-diffusion
system. Error bounds for arbitrary initial guesses are shown.}
\label{TableIV}
\begin{tabular}{lcccc}
Case& Data used & Noise level & $f$ & $k$ \\
\cline{1-5}
{\it(a)}&$u^{(1)}(x,t)$  & 0.00 & $0.0290 \pm 0.0001$ & $0.0535 \pm 0.0001$ \\
       &$u^{(2)}(x,t)$  & 0.02 & $0.0291 \pm 0.0001$ & $0.0537 \pm 0.0001$ \\
       &                & 0.05 & $0.0296 \pm 0.0003$ & $0.0540 \pm 0.0003$ \\
\cline{1-5}
{\it(b)}&$u^{(1)}(x,t)$  & 0.00 & $0.0291 \pm 0.0002$ & $0.0538 \pm 0.0003$ \\
       &                & 0.02 & $0.0296 \pm 0.0005$ & $0.0565 \pm 0.0003$ \\
       &                & 0.05 & $0.0303 \pm 0.0008$ & $0.0610 \pm 0.0003$ \\
\cline{1-5}
{\it(c)}& $u^{(1)}(x,t)$ & 0.00 & $0.0295 \pm 0.0002$ & $0.0540 \pm 0.0002$ \\
       & $0.25<x<0.75$  & 0.02 & $0.0307 \pm 0.0008$ & $0.0578 \pm 0.0002$ \\
       &                & 0.05 & $0.0321 \pm 0.0008$ & $0.0614 \pm 0.0003$
\end{tabular}
\end{table}


\begin{thebibliography}{xx}

\bibitem{Cr93} M. C. Cross and P. C. Hohenberg, Rev. Mod. Phys.
{\bf 65}, 851 (1993).

\bibitem{Gr89} P. Grassberger, Phys. Scr. {\bf 40}, 336 (1989).

\bibitem{Ba91} M. Bauer and W. Martienssen, J. Phys. A {\bf 24}, 4557 (1991).

\bibitem{Pa98} N. Parekh, V. Ravi Kumar, and B. D. Kulkarni,
Chaos {\bf 8}, 300 (1998);
N. Parekh, V. Ravi Kumar, and B. D. Kulkarni, Pramana {\bf 48}, 303 (1997).

\bibitem{St99a} R. Carreto-Gonz\'{a}lez, S. {\O}rstavik, J. Huke,
D. S. Broomhead, and J. Stark, Chaos {\bf 9}, 466 (1999);
S. {\O}rstavik,  R. Carreto-Gonz\'{a}lez and J. Stark, Physica D 
{\bf 147}, 204 (2000).

\bibitem{Si87} L. Sirovich,  Q. Appl. Math. {\bf{45}}, 561 (1987).

\bibitem{Ho96} P. Holmes, J. L. Lumley and G. Berkooz,
{\it Turbulence, Coherent Structures, Dynamical Systems and
Symmetry} (Cambridge University Press, Cambridge, 1996).

\bibitem{Ba83} H. Bock, {\it Progress in Scientific Computing},
eds. P. Deuflhard and E. Hairer, (Birkh\"{a}user, Boston 1983){\bf 2} 95; 
H. G. Bock and K. J. Plitt, {\it A Multiple Shooting Algorithm for
Direct Solution of Optimal Control Problems}, International federation
of automatic control, $9^{th}$ World congress, Budapest, (1984) (Pergamon,
Oxford, 1984).

\bibitem{Ti98} J. Timmer, Int. J. Bifurcation Chaos. {\bf 8}, 1505 (1998).

\bibitem{Ti00} J. Timmer, H. Rust, W. Horbelt, and H.U. Voss,
Phys. Lett. A {\bf 274}, 123 (2000).

\bibitem{Ba92} E. Baake, M. Baake, H. G. Bock, and K. M. Briggs, Phys. Rev.
A {\bf 45}, 5524 (1992).

\bibitem{Ba99} M. B\"{a}r, R. Hegger, and H. Kantz,
Phys. Rev. E {\bf 59}, 337 (1999).

\bibitem{Pa00} U. Parlitz and C. Merkwirth,
Phys. Rev. Lett. {\bf 84}, 1890 (2000).

\bibitem{Vo98} H. Voss, M. J. B\"{u}nner and M. Abel,
Phys. Rev. E {\bf 57} 2820 (1998);
H. Voss, in {\it Nonlinear Dynamics and Statistics},
Ed. A. Mees, (Birkha\"{u}ser, Boston, 2000).

\bibitem{Lo00} C. L\'{o}pez, A. \'{A}lvarez and E. Hern\'{a}ndez-Garc\'{i}a,
Phys. Rev. Lett. {\bf 85}, 2300 (2000).

\bibitem{Ka93} K. Kaneko ed., {\it Theory and Applications of Coupled Map
Lattices}, (John Wiley \& Sons Ltd., West Sussex, 1993);
J. P. Crutchfield and K. Kaneko, {Directions in Chaos}, (World
Scientific, Singapore 1987).

\bibitem{Wi95} F. H. Willeboordse and K. Kaneko, Physica D {\bf 86}, 428 (1995).

\bibitem{Me00} M. Meixner, S. M. Zoldi, S. Bose and E. Sch\"{o}ll,
Phys. Rev. E {\bf 61}, 1382 (2000).

\bibitem{Ma96} W. Mazin, K. E. Rasmussen, E. Mosekilde, P. Borckmans,
and G. Dewel, Maths. Comp. Simul. {\bf{40}}, 371 (1996).

\bibitem{Da92} I. Daubechies,
{\it Ten Lectures on Wavelets} (SIAM Publications, Philadelphia, 1992).

\bibitem{Wi99} R. W. Wittenberg and P. Holmes, Chaos {\bf 9}, 452 (1999).

\bibitem{Po78} M.J.D. Powell, {\it A Fast Algorithm for Nonlinearly Constrained Optimization Calculations} Lecture notes in Maths, {\bf 630}, (Springer,
Berlin, 1978).

\bibitem{Sc86}K. Schittkowski, Ann. Operation Res., {\bf 5}, 485 (1986).

\bibitem{Mc99} P. E. McSharry and L. A. Smith,
Phys. Rev. Lett. {\bf 83}, 4285 (1999).

\bibitem{Hi99} A. Hilgers and C. Beck, Europhys. Lett. {\bf 45}, 552 (1999).

\bibitem{Ka89} K. Kaneko, Physica D {\bf 34}, 1 (1989).

\bibitem{Zo97} S. M. Zoldi and H. S. Greenside,
Phys. Rev. Lett. {\bf 78}, 1687 (1997).

\bibitem{Ku84} Y. Kuramoto,  {\it {Chemical Oscillations,
Waves and Turbulence}}, (Springer, Berlin, 1984).

\bibitem{Fi85} R. J. Field and M. Burger,
{\it{Oscillations and Travelling Waves in Chemical Systems}},
(John Wiley, New York, 1985).

\bibitem{Gr90} P. Gray and S. K. Scott,
{\it {Chemical Oscillations and Instabilities}},
(Oxford University Press, 1990).

\bibitem{Se68} E. E. Selkov, Eur. J. Biochem. {\bf{4}}, 79 (1968).

\bibitem{Le94} K.J. Lee, W.D. McCormick, J.E. Pearson, and H.L. Swinney,
Nature {\bf 369}, 215 (1994).

\end{thebibliography}
\end{document}